# Extreme dissipation and intermittency in turbulence at very high Reynolds numbers

G. E. Elsinga[1], T. Ishihara[2] and J. C. R. Hunt[1,3]

[1]Laboratory for Aero and Hydrodynamics, Department of Mechanical, Maritime and Materials Engineering, Delft University of Technology, 2628CD Delft, The Netherlands
[2]Graduate School of Environmental and Life Science, Okayama University, Okayama 700-8530, Japan
[3]Department of Earth Sciences, University College London, London WC1E 6BT, United Kingdom

Extreme dissipation events in turbulent flows are rare, but they can be orders of magnitude stronger than the mean dissipation rate. Despite its importance in many small-scale physical processes, there is presently no accurate theory or model for predicting the extrema as a function of the Reynolds number. Here, we introduce a new model for the dissipation PDF based on the concept of significant shear layers, which are thin regions of elevated local mean dissipation. At very high Reynolds numbers these significant shear layers develop layered substructures. The flow domain is divided into the different layer regions and a background region, each with their own PDF of dissipation. The volume weighted regional PDFs are combined to obtain the overall PDF, which is subsequently used to determine the dissipation variance and maximum. The model yields Reynolds number scalings for the dissipation maximum and variance, which are in agreement with the available data. Moreover, the power law scaling exponent is found to increase gradually with the Reynolds number, which is also consistent with the data. The increasing exponent is shown to have profound implications for turbulence at atmospheric and astrophysical Reynolds numbers. The present results strongly suggest that intermittent significant shear layer structures are key to understanding and quantifying the dissipation extremes, and, more generally, extreme velocity gradients.

## 1. Introduction

Extreme viscous dissipation and the associated extreme strain rates are critical in many processes dominated by small-scale turbulence. For example, these events can lead to local flame extinction and pollutant formation [1], affect the onset of detonation in supernovae [2], and are sites for droplet collision and growth in clouds [3,4]. Furthermore, intense dissipation enhances the intermittency of the scalar dissipation rate and can be a source of local heating, especially at higher Reynolds numbers, which is relevant for quantifying mixing and chemical reaction rates [5-7]. Extreme dissipation also sets the smallest scale in a turbulent flow, thereby posing resolution requirements for experiments and numerical simulations [8,9], which is particularly relevant for studies into the mentioned small-scale processes.

The scaling of extreme dissipation requires further study, especially for extrapolating results from laboratory experiments and direct numerical simulations (DNSs) to applications in industry, the atmosphere and astrophysics at much higher Reynolds numbers. The difficulties arise due to the highly complex and intermittent nature of the dissipation rate of kinetic energy, $\varepsilon = 2\nu S_{ij}S_{ij}$. Here, $\nu$ is the kinematic viscosity and $S_{ij}$ is the strain-rate tensor, which is the symmetric part of the velocity gradient tensor. On the one hand, its mean, $\langle\varepsilon\rangle$, appears constant in (near) equilibrium turbulence at sufficiently large Reynolds number when normalized by large-scale quantities $U^3/L$, where $U$ is the root-mean-square of the velocity fluctuations and $L$ is the integral length scale [10,11]. However, the value for the normalized mean dissipation, $D = \langle \varepsilon \rangle L/U^3$, depends on the flow and the type of forcing [12]. This can be understood as an energy balance between the dissipation rate, $\langle \varepsilon \rangle$, and the turbulent kinetic energy production rate at large scales, where the latter must be equal to $DU^3/L$ in order to achieve equilibrium. Therefore, the ratio $D$ is more informative of production rather than dissipation. Indeed, when the large scales are out-of-equilibrium and evolve structurally, $D$ is no longer a constant and depends on the local flow conditions [13,14]. On the other hand, extreme dissipation and the higher order statistical moments of $\varepsilon$ are observed to increase with Reynolds number when normalized by the Kolmogorov scales, or equivalently by $\langle \varepsilon \rangle$ [9,15]. Hence, the probability density function (PDF) of the dissipation rate remains strongly Reynolds number dependent. This is closely associated with small-scale intermittency increasing with Reynolds number [16-18]. The development of the dissipation PDF with Reynolds number is also evident from the dissipation variance. For homogeneous isotropic turbulence, the dissipation variance is directly related to the flatness factor of the longitudinal velocity gradient [19,20], which increases with Reynolds number [15,17,20].

Simple Reynolds number scaling laws may be obtained from fits to the data. Donzis, Yeung & Sreenivasan [21] fitted a stretched exponential to the tail of the dissipation PDF at four different $Re_\lambda$ between 140 and 650, where $Re_\lambda$ is the Reynolds number based on the Taylor micro-scale. Their results show that the tails become wider with increasing $Re_\lambda$. The $Re_\lambda$ dependence was further quantified by Buaria et al. [9], who fitted a power law to the coefficients of the stretched exponential over the same $Re_\lambda$ range. They also suggested that a constant power law may not be sufficient and that the power itself may be Reynolds number depended without specifying. Indeed, Elsinga et al. [22] have shown that the velocity increment associated with extreme dissipation follows a power law scaling with $Re_\lambda$, and that the exponent changes around $Re_\lambda = 250$ (their figure 19). This transition is related to the development of the vorticity stretching motions [22]. Because such transitions are not accounted for, the constant power laws apply only as an approximation over a limited range of Reynolds numbers.

Similarly, theories have predicted constant power laws. For example, in the (multifractal) descriptions [23-25], the smallest scale needed to capture the higher order dissipation moments, hence extreme dissipation, scales as $LRe_\lambda^{-2}$, which corresponds to $\varepsilon_{max} \sim \langle \varepsilon \rangle Re_\lambda^2$. Not only is the exponent constant, also its value is inconsistent with the data, which reveal that the exponent is closer to 1.5 at $Re_\lambda \sim 400$ ([9], see also §4). Alternatively, Kolmogorov [26] proposed a lognormal distribution of $\varepsilon$, which leads to $\varepsilon_{max} = \langle \varepsilon \rangle Re_\lambda^\beta$ with $\beta \leq 1$ decreasing with the Reynolds number (see §4). The exponent and the decreasing trend, however, are inconsistent with the data (§4).

Presently, there is no suitable theory to explain the observed Reynolds number dependence of the dissipation PDF, its variance and also its extremes. Any successful theory, it seems, will need to account properly for the intermittency, which in turn may be related to structure development. Specifically, elevated levels of the dissipation rate have been associated with significant shear layer structures, which are thin, intermittent structures of high shear, bounding the large-scale flow regions ([27,28], §2(a)). The regions outside these shear layers contain low, background level, dissipation. The significant shear layer structures were studied in DNS of homogenous isotropic turbulence up to $Re_\lambda = 1131$. However, large-scale shear layers have also been observed at much higher Reynolds numbers in the earth's atmosphere [29] and in molecular clouds [7,30]. See figure 7(right) in [7] for a striking example of a long and thin structure of elevated velocity shear, which extends over $\sim 2 \cdot 10^{16}$ m in a molecular cloud. Therefore, the present analysis based on the concept of significant shear layers is considered relevant all the way up to the highest Reynolds numbers. High Reynolds number geophysical and astrophysical flows can, however, include additional effects of buoyancy and compressibility, among other things, which are not considered here. Nevertheless, incompressible turbulence theory finds important use under these conditions in order to understand the main effects of turbulence.

Here, we develop a model for the PDF of the dissipation rate based on the concept of significant shear layers. Our aim is to generate an advanced understanding of the Reynolds number dependence in the dissipation PDF. The first results show that the prediction of the dissipation variance and extreme dissipation are consistent with the available data for homogeneous isotropic turbulence. Furthermore, the implications for flows at geophysical and astrophysical Reynolds numbers are considered.

The structure of the paper is as follows. The Reynolds number scaling of the volume occupied by the significant shear layers is discussed first in §2. Also, the breakdown of these layers into sublayers at very high Reynolds numbers is considered (§2(b)). Then, dissipation distributions are associated with the layers and the non-layer regions of space, which are combined to yield the overall model PDF (§3). From the model PDF, the dissipation variance and extremes are determined and compared to available data (§4). Additionally, some implications for the temporal and spatial resolution in experiments and DNS are discussed briefly (§5). The findings are summarized in §6.

## 2. Intermittency and flow structure

### (a) Significant shear layers

The significant shear layers have been observed in snapshots of homogenous isotropic turbulence at high Reynolds number [28] as well as in the statistics of the turbulent strain field [22]. Recent visualizations suggest they may be the arms of large-scale spiral vortices, but their exact origin and dynamics are not important here. Discussions of shear layer dynamics can be found in [31,32]. Alternatively, such structure formation may be understood from weak solutions of the 3D Navier-Stokes equations and dimensional analysis [33]. However, for the present purpose, only the kinematic scaling matters. Significant shear layers separate large-scale regions of approximately uniform flow (i.e. low-level velocity fluctuations), which determine the velocity difference, and hence the shear, across the layer. This velocity difference is of order $U$. The thickness of the significant shear layer is approximately $4\lambda_T$ [22,28], where $\lambda_T = \sqrt{\frac{5\nu\langle u'_i u'_i\rangle}{\langle \varepsilon \rangle}}$ is the Taylor micro-scale and $u'_i$ are the velocity fluctuations. Since these layers continue along the edges of the large-scale regions, the length of the layer is order $L$ in the other directions. This means that the volume occupied by the significant shear layer scales according to $\lambda_T L^2$. An illustration of this scaling is given in figure 7 of Hunt et al. [32]. When normalized by the full flow domain ($\sim L^3$), the volume fraction of the layers scales according to $\lambda_T L^{-1} \sim Re_\lambda^{-1}$. As the Reynolds number, hence the scale separation, increases, the volume fraction of the layers decreases.

The large velocity difference across this thin layer allows for the development of very intense small-scale structures within the layer. These small-scale structures include extreme dissipation events. Ishihara, Kaneda & Hunt [28] have shown that the average dissipation rate within the layer can be more than ten times the mean dissipation rate at $Re_\lambda = 1131$. Furthermore, a visualization of the dissipation rate at $Re_\lambda = 675$ reveals layer-like clusters of extreme dissipation (figure 4 in [11]), which is consistent with the concept of significant shear layers. Statistical support for extreme dissipation within the significant shear layers is provided by the average velocity field conditioned on extreme dissipation. It reveals that wherever extreme dissipation occurs, it is bounded by energetic large-scale motions, whose strength is of order $U$ as expected for significant shear layers [22]. Therefore, it is concluded that significant shear layers are important in explaining the intermittency of the dissipation rate (and of small-scales in general). This is examined in more detail in §§3 and 4.

For the significant shear layers to be fully developed, a minimum $Re_\lambda$ of 250 is required. Below this threshold, the layer thickness is smaller than the small-scale coherence length [22], which means that the small scales cannot properly fit inside the layer. Note that 'underdeveloped' shear layers are observed at lower Reynolds numbers. For example, at $Re_\lambda =$ 170 the small-scale vortices are seen to cluster at the edges of the large-scale flow regions (figure 14 in [34]). However, the scaling of these underdeveloped layers will be slightly different from those at $Re_\lambda > 250$. Below $Re_\lambda = 45$, the large and small-scale coherence lengths are identical [22], and layers cannot be distinguished from the large-scale structure.

Generally, the large scales are flow dependent in terms of their geometry and dynamics (e.g. large-scale bursting behaviour [35]), so it is reasonable to assume that their edges, i.e. the significant shear layers, are also flow dependent. However, qualitative similarity in the layer structure has been noted across different flows [22,34], which suggests that the scaling laws developed herein apply broadly. However, the model coefficients may be flow dependent (see also §2(b)).

## (b) Development of substructures

Next, we consider the possibility that the significant shear layers develop substructures, which would increase intermittency even more. At high Reynolds number, the significant shear layers are fully turbulent and contain small-scale structures (e.g. vortices). At even higher Reynolds number, the layer region is so large with respect to the small-scales, that it can be considered as an independent turbulent (shear) flow region with its own associated local Reynolds number $Re_\lambda^*$. Then from previous investigations, we know that, if the Reynolds number is sufficiently large, significant shear layers appear. In this case, these layers appear within the original significant shear layer as a substructure. This will be referred to as the breakdown into substructure. As shown below, the breakdown into substructure would require very high $Re_\lambda$, well in excess of 1000. Some observational support for the existence of substructure, i.e. sublayers, is discussed in §4.

The local $Re_\lambda^*$ is established based on the range of scales within the layer. For the large scales within the significant shear layer, we require that the local integral length scale $L^*$ of the layer turbulence satisfies $4L^* = 4\lambda_T$ such that four local integral scales are contained within the layer thickness. This criterion is chosen to avoid confinement effects, meaning that $L^*$ is the largest scale not directly affected by the layer thickness. For similar reasons, the computational domains in DNS are usually larger than four integral length scales (e.g. [36]).

To estimate the local Kolmogorov scale $\eta^*$ of the layer turbulence, we use the dissipation ratio $\varepsilon^*/\langle\varepsilon\rangle$, where $\varepsilon^*$ is the mean dissipation rate averaged over the shear layer volume $V^*$ and $\langle\varepsilon\rangle$ is the mean dissipation rate averaged over the entire flow volume $V$. As discussed in §2(a), the volumes are related according to $V^*/V = \alpha^{-1} Re_\lambda^{-1}$, where $\alpha$ is a constant. Furthermore, a low-level background dissipation rate, $\varepsilon_{bg}$, is defined, which is constant throughout the entire volume. This yields $\langle\varepsilon\rangle V = (\varepsilon^* - \varepsilon_{bg})V^* + \varepsilon_{bg}V$. Assuming that the background dissipation rate is a fixed percentage of the mean, i.e. $\varepsilon_{bg} = b\langle\varepsilon\rangle$, it follows that:

$$\varepsilon^* = \langle\varepsilon\rangle[b + (1-b)\alpha Re_\lambda] \qquad (2.1)$$

Using the definition of the Kolmogorov length scale, we obtain:

$$\frac{\eta^*}{\eta} = \left(\frac{\langle\varepsilon\rangle}{\varepsilon^*}\right)^{\frac{1}{4}} = [b + (1-b)\alpha Re_\lambda]^{-\frac{1}{4}} \qquad (2.2)$$

where $\eta = \left(\frac{\nu^3}{\langle\varepsilon\rangle}\right)^{\frac{1}{4}}$ is the global Kolmogorov micro-scale based on the dissipation rate averaged over the entire flow volume.

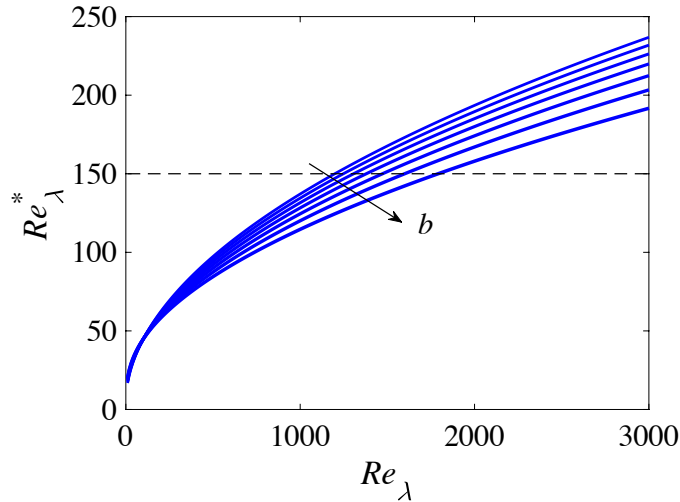

**Figure 1.** The local Reynolds number for the turbulence within the significant shear layer, $Re_\lambda^*$, versus the global Reynolds number, $Re_\lambda$ (equation 2.4). Results are for $\alpha = 0.010$ and for different $b$ between 0.2 and 0.8 in increments of 0.1.

Using the local length scales $\eta^*$ and $L^*$, it is possible to relate the Reynolds numbers of the layer turbulence $Re_\lambda^*$ to that of the entire flow $Re_\lambda$. Using the definition $Re_\lambda = \frac{1}{\sqrt{15}}\left(\frac{\lambda_T}{\eta}\right)^2$, the ratio of layer scales is given by:

$$\frac{L^*}{\eta^*} = \frac{\lambda_T}{\eta}[b + (1-b)\alpha Re_\lambda]^{\frac{1}{4}} = 15^{1/4} Re_\lambda^{\frac{1}{2}}[b + (1-b)\alpha Re_\lambda]^{\frac{1}{4}} \qquad (2.3)$$

which defines a Reynolds number according to the relation $\frac{L}{\eta} \approx 15^{-\frac{3}{4}} D Re_\lambda^{\frac{3}{2}}$. The normalized mean dissipation rate, $D$, is approximately equal to 0.5 for the present case of forced homogenous isotropic turbulence and $Re_\lambda > 100$ [37]. Inserting $\frac{L^*}{\eta^*} \approx 15^{-\frac{3}{4}} D Re_\lambda^{*\frac{3}{2}}$ into equation (2.3) and rearranging terms yields:

$$Re_\lambda^* = 15^{2/3} D^{-2/3} Re_\lambda^{\frac{1}{3}}[b + (1-b)\alpha Re_\lambda]^{\frac{1}{6}} \qquad (2.4)$$

This relation is shown in figure 1. Here and throughout the paper, $\alpha$ is taken to be 0.010, which is consistent with the volume ratio resulting from $4\lambda_T$ thick layers bounding $L$ wide large-scale flow regions, as expected for significant shear layers (§2(a)). Further note that $b = 0.67$ is used in the remainder of the paper, because it yields a good correspondence with the data in §4. Moreover, $b = 0.67$ results in $\varepsilon^*/\varepsilon_{bg} \approx 6.4$ at $Re_\lambda = 1100$ (equation 2.1), which is consistent with the ratio of the average dissipation inside and outside the significant shear layer as reported by Ishihara et al. [28]. However, the coefficients $\alpha$ and $b$ may still be flow dependent, similar to $D$. The present numerical values are for incompressible homogenous isotropic turbulence. We emphasize that the power law exponents are independent of $\alpha$, $b$ and $D$ (see equations 2.1-2.5). Nevertheless, the transitional Reynolds numbers and the dissipation magnitude dependent on these coefficients.

In a previous study, we found $Re_\lambda \approx 250$ marking the onset of fully developed significant shear layers within the turbulent flow [22]. However, as mentioned, underdeveloped shear layers are already seen at $Re_\lambda = 170$. To allow for a gradual transition, the threshold for the first appearance of significant shear layers is relaxed to $Re_\lambda = 150$. Similarly, we expect that beyond $Re_\lambda^* = 150$ the first signs of substructures are present within the significant shear layer. These substructures are again of shear layer type. Therefore, they are referred to as sublayers. Using $\alpha = 0.010$ and $b = 0.67$ in equation (2.4), $Re_\lambda^* = 150$ corresponds to $Re_\lambda = 1560$ for the full flow, which is achievable by DNS [38]. Then, the volume fraction occupied by sublayers is given by:

$$\frac{V_{sublayer}}{V} = \frac{V_{sublayer}}{V^*}\frac{V^*}{V} = \alpha^{-2} Re_\lambda^{*-1} Re_\lambda^{-1} \qquad (2.5)$$

The process of generating substructure repeats, that is, the sublayer can develop sub-sublayers. This is expected to occur when $Re_\lambda^* = 1560$, by analogy with the above, which corresponds to $Re_\lambda = 1.8 \cdot 10^5$. In principle, this process can be repeated endlessly (a next level substructure appears at $Re_\lambda = 2 \cdot 10^9$), but we terminate at $Re_\lambda = 10^8$ given the fact that $10^7$ is reached only in molecular clouds with active internal star formation [30] and in supernovae [2]. The development of sublayers and sub-sublayers is illustrated in figure 2.

The resulting volume fractions of the significant shear layers, its sublayers and sub-sublayers are shown in figure 3. The volume fractions follow a power law scaling with $Re_\lambda$, where the exponents are -1, -3/2 and -7/4 respectively. Therefore, the volume occupied by these layers decreases very quickly with Reynolds number, hence intermittency increases.

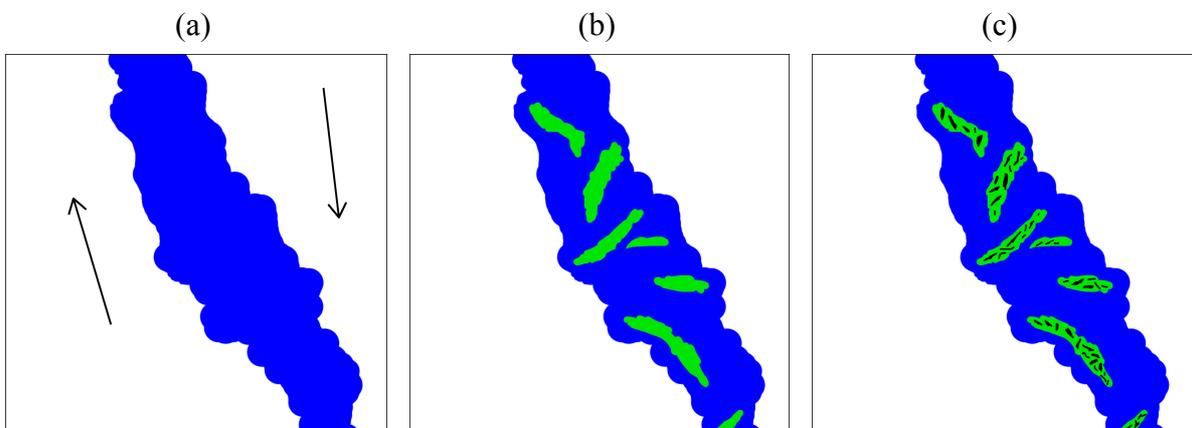

**Figure 2.** (a) Significant shear layer structure (blue) within a part of the flow domain ($Re_\lambda > 150$). (b) Significant shear layer structure with sublayers in green ($Re_\lambda > 1560$). (c) Significant shear layer structure with sub-sublayers in black ($Re_\lambda > 1.8 \cdot 10^5$). For illustration purposes only; layers are not to scale.

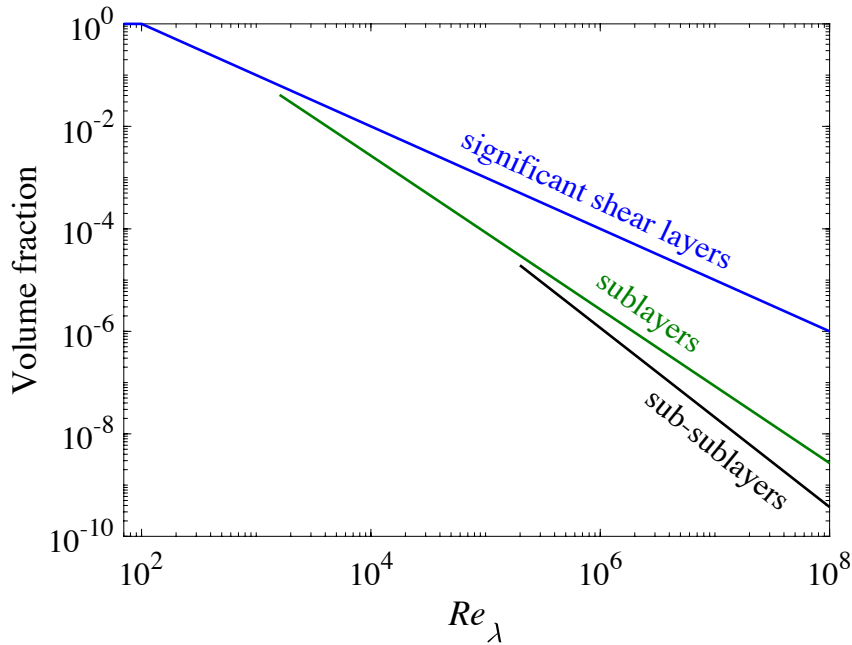

**Figure 3.** The volume fraction of the significant shear layer (blue), its sublayers (green) and sub-sublayers (black). The volume fraction is defined relative to the entire flow volume. Results are for $\alpha = 0.010$ and $b = 0.67$.

The above breakdown of the significant shear layer is different from the classic shear layer transition due to Kelvin-Helmholtz instabilities. In the latter, the shear layer breaks up into a series of vortex tubes (e.g. [20]), which has been observed in DNS of homogenous isotropic turbulence [39-41]. However, this concerns small-scale shear layers, which develop small-scale vortices. The scale of the structure does not change. In that sense, there is no development of a smaller-scale substructure. Therefore, it is very different from the process described here, where the $\lambda_T$-scale shear layer is already turbulent and contains smaller-scale substructures that further develop at higher Reynolds number. Note that for $Re_\lambda > 150$, the Reynolds number of the layer is $Re_\lambda^* > 50$ (figure 1) indicating that the layer can be turbulent. At low Reynolds number, these substructures are simply the $\eta$-scale structures, i.e. intense vortices and dissipation sheets, while at high Reynolds number a much richer, fully turbulent substructure emerges consisting of large scale regions (relative to the layer thickness) bounded by significant sublayers containing the intense vortices and dissipation.

Furthermore, we comment that the significant shear layers and the sublayer structures have disparate time scales. Since the significant shear layers result from large-scale structures rubbing against each other, their lifetime is of the order of $L/U$. Similarly, the characteristic time scale for the sublayers is $L^*/U = \lambda_T/U$, which is more than 1.5 orders of magnitude smaller than $L/U$ at the Reynolds number where these sublayers first appear. Note that the velocity scale for the internal structures of the significant shear layer is $U$, which is supported by observations [22,28] and is consistent with the assumption that the significant shear layers' relative contribution to the overall dissipation rate is independent of the Reynolds number. The relatively slow evolution of the significant shear layer results in a quasi-stationary forcing of its internal turbulence, which allows the internal turbulence to be fully developed. This justifies our kinematic approach.

In the present model, it is implicitly assumed that the large-scale regions bounding the significant shear layers do not develop substructures at large Reynolds number. This is equivalent to assuming that the dissipation in these regions is Reynolds number independent and can be regarded as background dissipation. The breakdown of the large scales into substructure would have introduced an inconsistency, since it implies that the largest flow scale becomes considerably smaller, thereby reducing scale separation at increasing Reynolds number. It is clear that more research is needed on the flow structures within these large-scale regions. However, the goal is to predict the dissipation extremes, which are associated with the significant shear layers. What happens to the large scales at high Reynolds number is of secondary importance.

## 3. Dissipation distribution

In this section, the PDF of the dissipation rate is estimated based on the structural evolution discussed in §2. However, first the local mean dissipation rate needs to be determined for each of the different flow regions.

At low Reynolds number ($Re_\lambda < 150$), there is just one flow region. Shear layers and the large-scale regions are inseparable and equivalent. Therefore, the local mean is equal to the global mean $\langle \varepsilon \rangle$. When $Re_\lambda > 150$, the significant shear layers separate from the large-scales and the mean dissipation rate within these layers, $\varepsilon^*$, is given by equation (2.1). This is shown in figure 4 by the blue solid line. For the regions outside the layer, the local mean dissipation rate is given by the background dissipation rate $\varepsilon_{bg} = b\langle \varepsilon \rangle$ (gray dashed line in figure 4).

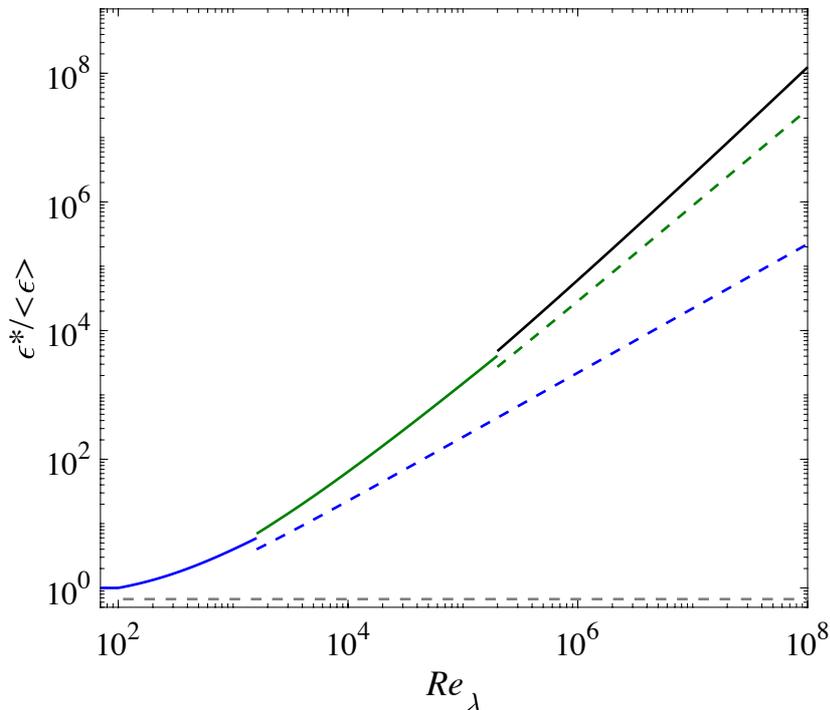

**Figure 4.** The local mean layer dissipation rate within the significant shear layer (solid-blue), its sublayers (solid-green) and sub-sublayers (solid-black). Dashed lines show the background dissipation in the corresponding regions when a substructure emerges. Results are for $\alpha = 0.010$ and $b = 0.67$.

When the sublayers emerge within the significant shear layer ($Re_\lambda > 1560$), the mean dissipation rate over the entire significant shear layer is still given by $\varepsilon^*$ (equation (2.1)), while the regions outside these layers are characterized by $\varepsilon_{bg}$. However, $\varepsilon^*$ needs to be subdivided over the sublayers (green in figure 2(b)) and the regions outside the sublayers but within the significant shear layer (blue in figure 2(b)). The latter are considered as the local background to the sublayers. Therefore, their mean local dissipation is given by $b\varepsilon^*$, which is shown by the dashed blue line in figure 4. By analogy with equation (2.1), the local mean dissipation rate in the sublayers is given by:

$$\varepsilon^*_{sublayer} = \varepsilon^*[b + (1-b)\alpha Re_\lambda^*] \tag{3.1}$$

where $Re_\lambda^*$ is the Reynolds number for the significant shear layer turbulence (equation 2.4). The sublayer's mean dissipation rate is shown by the green solid line in figure 4.

The process repeats when the sub-sublayers appear ($Re_\lambda > 1.8 \cdot 10^5$). For the sub-sublayer's background (green in figure 2(c)), the mean dissipation rate is $b\varepsilon^*_{sublayer}$, which is shown by the dashed green line in figure 4. The sub-sublayer's mean dissipation rate ($\varepsilon^*_{sub-sublayer}$, black solid line in figure 4) is obtained using the Reynolds number for the sublayer turbulence and $\varepsilon^*_{sublayer}$ on the right-hand-side of equation (3.1). The mean dissipation rate in the sublayer's background (blue in figure 2(c)) and the region outside the significant shear layers are still given by $b\varepsilon^*$ and $\varepsilon_{bg}$ respectively (dashed blue and gray lines in figure 4).

Then the dissipation rate in each region is assumed to be lognormally distributed according to:

$$P(\varepsilon/\langle\varepsilon\rangle) = \frac{\langle\varepsilon\rangle}{\varepsilon\sqrt{2\pi}\sigma} \exp\left(-\frac{(\ln(\varepsilon/\langle\varepsilon\rangle) - \mu)^2}{2\sigma^2}\right) \tag{3.2}$$

Its mean, i.e. $\exp(\mu + \sigma^2/2)$, corresponds to the local mean dissipation rate of the given region, as determined above and shown in figure 4. The parameter $\sigma$ is a measure for the width of the distribution on a log scale, which is taken constant at $\sigma = 1$. The latter is motivated by the fact that the PDFs of the enstrophy inside and outside the significant shear layer are similar in shape and width (figure 8 in [28]). The main difference appears to be explained by the change in the local mean enstrophy. The dissipation rate is expected to behave similarly, since both dissipation rate and enstrophy are small-scale quantities that depend on the velocity gradients squared. The lognormal distribution is an approximation, which does not accurately describe the probability at low dissipation rate. Also, the high dissipation tail is somewhat better approximated by a stretched exponential with an exponent around 0.25 [8,21]. These aspects could be improved in the future.

The PDFs of the individual regions are weighted according to their volume fraction (figure 3) and summed to produce the overall PDF. Examples of the resulting overall PDFs are shown in figure 5 for different Reynolds numbers (black solid lines). Furthermore, they are compared with basic lognormal distributions, which represent a model without the contribution from significant shear layers (dashed lines). The comparison is completed by data from actual DNSs of homogenous isotropic turbulence (red lines). Here, the values of $\varepsilon$ are calculated using a Fourier spectral method as used in the DNSs. The details of the DNS data are provided in table 1.

**Table 1.** DNS parameters and main turbulence characteristics. The simulations were performed on a $2\pi$ sized domain using the code described in [15,37,38]. The original source of each dataset is given in the last column. Here (2007) refers to Ishihara et al. [15], while (Ext-2007, -2016) refer the simulations presented in [15] and [38] respectively, which have been extended in terms of increased spatial resolution and increased integration time. Furthermore, $N_{grid}$ is the number of grid points in each direction, $k_{max}$ is the maximum wavenumber retained in the DNS, $T=L/U$ is the eddy turnover time, $t_F$ is the total integration time, and $t_E$ is the integration time after increasing spatial resolution without changing viscosity. $(3/2)U^2 = 0.5$ for all the runs.

| Run | $N_{grid}$ | $Re_\lambda$ | $k_{max}\eta$ | $10^4\nu$ | $\langle\varepsilon\rangle$ | $L$ | $\lambda_T$ | $10^3\eta$ | $T$ | $t_F/T$ ($t_E/T$) | source |
|---|---|---|---|---|---|---|---|---|---|---|---|
| 256-2 | 256 | 94 | 2 | 20 | 0.0936 | 1.10 | 0.327 | 17.1 | 1.91 | 5.2 | 2007 |
| 512-2 | 512 | 173 | 2 | 7 | 0.0795 | 1.21 | 0.210 | 8.10 | 2.10 | 4.8 | 2007 |
| 1024-2 | 1024 | 268 | 2 | 2.8 | 0.0829 | 1.12 | 0.130 | 4.03 | 1.94 | 5.2 | 2007 |
| 2048-2 | 2048 | 446 | 2 | 1.1 | 0.0762 | 1.11 | 0.0850 | 2.04 | 1.92 | 4.4 | 2007 |
| 4096-2 | 4096 | 730 | 2 | 0.44 | 0.0711 | 1.12 | 0.0556 | 1.05 | 1.94 | 6.8 (1.6) | Ext-2016 |
| 8192-2 | 8192 | 1101 | 2 | 0.173 | 0.0794 | 1.10 | 0.0330 | 0.505 | 1.91 | 6.3 (1.0) | Ext-2016 |
| 2048-4 | 2048 | 272 | 4 | 2.8 | 0.0804 | 1.11 | 0.132 | 4.06 | 1.92 | 5.7 (0.5) | Ext-2007 |
| 8192-4 | 8192 | 739 | 4 | 0.44 | 0.0693 | 1.23 | 0.0564 | 1.05 | 2.13 | 5.3 (0.6) | Ext-2016 |

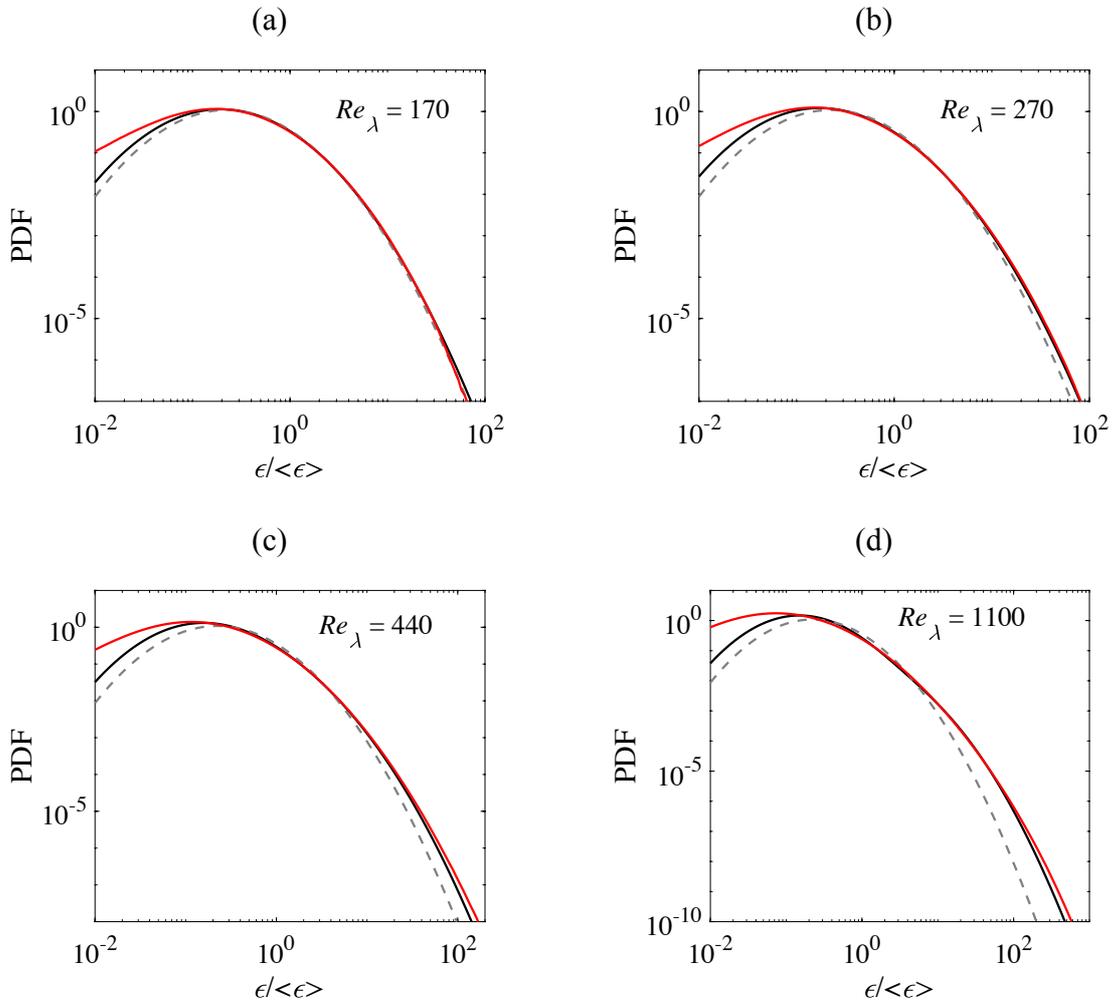

**Figure 5.** PDFs of the dissipation rate, comparing the model prediction (black) with a lognormal distribution (gray dashed lines) and DNS data at the closest available $Re_\lambda$ and $k_{max}\eta = 2$ (red). The model results are for $\alpha = 0.010$ and $b = 0.67$. DNS data sources are given in table 1.

At the lowest Reynolds number in figure 5 ($Re_\lambda = 170$), the significant shear layers have barely developed. Consequently, the model is very close to a lognormal distribution. However, at higher Reynolds numbers ($Re_\lambda > 250$), the model is clearly different from the lognormal distribution due to the significant shear layer contributions. When compared to the DNS, the model seems to accurately capture the departure from the lognormal distribution and the widening of the high dissipation tail with increasing $Re_\lambda$. It suggests that the development of extreme dissipation can be understood from the significant shear layers.

On the low dissipation side of the PDFs, the agreement with the DNS is not very good. This disagreement persists across all Reynolds numbers. As already mentioned, the lognormal distribution is not suitable to describe low dissipation events. However, our main interest is in the high dissipation tail and extreme dissipation, for which the lognormal distribution is appropriate as evident from figure 5(a).

Figure 6 presents the model PDFs of the dissipation rate over an extended range of Reynolds numbers. Furthermore, the PDFs are multiplied by $N^3$, where $N = \frac{5L}{3\eta}$. The result can be interpreted as the histogram of the dissipation rate for a $5L$ flow domain, typical for a DNS, sampled at $3\eta$, i.e. a representative size for the small-scale structure, when the bin size is equal to $\langle\varepsilon\rangle$. Note, that the specific choices for the prefactors do not affect the Reynolds number scaling behavior, but $N \propto \frac{L}{\eta}$ is important to account for the increase in small-scale detail at increasing Reynolds number. Hence, $N^3$PDF = 1 can be understood as occurring only once in the domain (on average). The DNS data used for comparison have not converged down to this level. Therefore, in §4, $N^3$PDF = 100 is used as a more robust criterion to define the maximum dissipation rate.

The results in figure 6 clearly indicate that the intermittency increases with $Re_\lambda$, in accordance with expectation. The high dissipation rate tails deviate strongly from the lognormal (dashed lines), which is due to the decreasing volume fraction of the layers (figure 3) and the corresponding increase in the local mean dissipation rate within these layers (figure 4).

For $Re_\lambda > 10^6$, a local minimum in the PDF is seen to the right of the main lobe. This is likely to be an artifact related to the assumption that all significant shear layers have exactly the same local mean dissipation rate $\varepsilon^*$ (equation (2.1)). It is reasonable to assume that $\varepsilon^*$ itself is distributed, which would smoothen the transition from the main lobe (corresponding to the background) to the tail (determined by the dissipation in the layers). These aspects are left for future studies. Here, the expected smooth distribution is simply illustrated by the dashed red line in figure 6 for $Re_\lambda = 10^7$.

Before proceeding, we briefly comment on the similarities and differences between the proposed model for the dissipation PDF and the well-known ideas of Kolmogorov [26] and Castaing et al. [42]. In their approaches, the regional variations in turbulence strengths are continuous and described by a lognormal distribution of the (local average) dissipation rate, whereas in our case we are dealing with a finite number of discrete flow regions (i.e. layer, sublayers and sub-sublayers) of specific average strengths. Because we convolve each regional average with a lognormal to account for some variation, we end up with the same lognormal distribution as [26,42] in case there is only a single flow region. However, when additional regions appear at higher Reynolds numbers, the low-probability tail of the overall dissipation PDF is modified by the contributions from the different layers. The effect is confined to the low-probability tail, because the volume fraction occupied by the layers is low. The high-probability main lobe is still dominated by the lognormal distribution associated with the large-scale background region (figure 6). Therefore, the differences with respect to Kolmogorov [26] and Castaing et al. [42] appear only in the low-probability high-dissipation tail (PDF < $10^{-5}$), which is usually not considered when fitting the Castaing et al. PDF shape to actual data. It is important to note that the width of their PDF is a fitting parameter. The Castaing et al. model is unable to predict the Reynolds number development of the width and the dissipation rate extrema, as we do in §4. Moreover, their PDF is intended to describe the energy transfer rate in the inertial range, where the intermittency and the PDF tails are less pronounced as compared to the resolved dissipation rate.

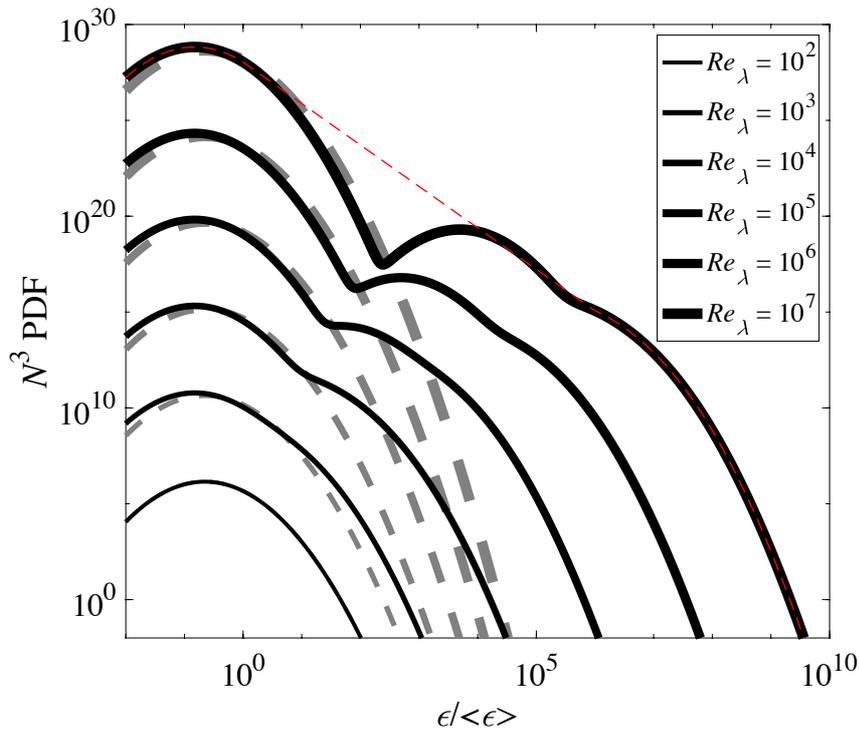

**Figure 6.** Scaled PDFs of the dissipation rate for different Reynolds numbers, where $N = \frac{5L}{3\eta}$. Black lines show model results using $\alpha = 0.010$ and $b = 0.67$. The dashed grey lines depict a lognormal distribution for the corresponding Reynolds numbers. The dashed red line shows the model PDF for $Re_\lambda = 10^7$ with the local minima removed by hand.

## 4. Extreme dissipation and dissipation moments

The maximum dissipation rate, $\varepsilon_{max}$, is deduced as the dissipation rate corresponding to $N^3$PDF = 100 (see also §3). This measure is considered to be more robust as compared to simply taking the maximum dissipation rate in the instantaneous flow fields, which fluctuates considerably and may include extraneous effects related to the numerical details of the simulation (or a measurement).

The predicted maximum dissipation rate increases quickly with Reynolds number (figure 7, black solid line). It follows a similar trend as the mean dissipation rate in the layer (figure 4), which is implied by the constant width of the lognormal distribution for the dissipation rate within the layer (§3). As noted in the introduction, the maximum dissipation does not follow a constant power law. Rather the exponent gradually changes from 1 initially to approximately 7/4 at the end of the Reynolds number range considered. The DNS data is consistent with the model predictions up to the highest, currently available Reynolds number ($Re_\lambda = 1100$). Note a slight underestimation of $\varepsilon_{max}$ (-15%) by the model in the $Re_\lambda$ range 400–1100, but the local slope, corresponding to an exponent of 1.3, is accurate to within 0.05. Furthermore, increasing the spatial resolution of the DNS to $k_{\max}\eta = 4$ has virtually no effect on the maximum dissipation rate in the present range (compare blue and green symbols in figure 7). Hence, $k_{\max}\eta = 2$ is considered sufficient. Lower resolution data at $k_{\max}\eta = 1$ was qualitatively consistent in terms of the $Re_\lambda$ dependence, but showed a considerable decrease of the maximum value (not shown).

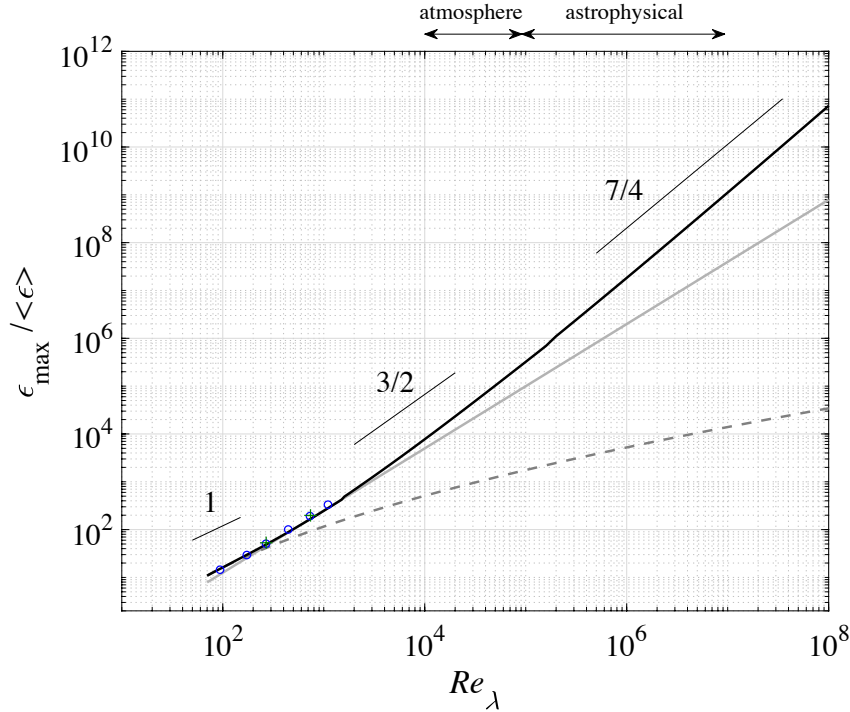

**Figure 7.** Maximum dissipation rate, $\varepsilon_{max}$, predicted by the present model (black solid line) and a lognormal distribution (grey dashed line). The model results are for $\alpha = 0.010$ and $b = 0.67$. Blue (o) and green (+) symbols indicate the maximum dissipation rate obtained from DNS data of homogenous isotropic turbulence at resolution $k_{\max}\eta = 2$ and 4 respectively (data sources are given in table 1). Power laws with exponents 1, 3/2 and 7/4 are indicated for reference. The grey solid line shows a constant power law with exponent 1.3, which corresponds to the model's and the DNS's scaling exponent in the range $400 < Re_\lambda < 1100$. The differences in $\varepsilon_{max}$ between a constant power law extrapolation (grey line) and the model (black line) are significant at the very high Reynolds numbers associated with atmospheric and astrophysical turbulence.

A $Re_\lambda^2$ scaling range, as predicted by multifractal theories [24], is not observed up to at least $Re_\lambda = 10^8$. Alternatively, the largest velocity gradients in the flow can be estimated to scale with $U$ and $\eta$, which results in $\frac{\varepsilon_{max}}{\langle\varepsilon\rangle} \sim \left(\frac{U}{\eta}\right)^2 \left(\frac{u_\eta}{\eta}\right)^{-2} \sim Re_\lambda^1$. This power scaling is observed only for $Re_\lambda$ around 100 (figure 7), which is consistent with the $Re_\lambda^{1/2}$ scaling of the velocity gradient in intense vortices at low $Re_\lambda$ (e.g. [43]). Clearly, $Re_\lambda^1$ scaling is insufficient to explain the Reynolds number trend beyond $Re_\lambda = 250$, where the volume occupied by intense dissipation, i.e. intermittency, plays an important role. Basically, it changes the local mean dissipation rate, thereby reducing the smallest length scale locally, see §5. The maximum energy, hence $U$, cannot change. A $Re_\lambda^{1.5}$ scaling for extreme dissipation has been observed by Buaria et al. [9]

for $Re_\lambda$ up to 650, whereas we find an exponent of 1.3 in this range. The small difference may be explained by the different measures used to quantify extreme dissipation. A $Re_\lambda^{1.5}$ scaling is, however, returned by the present model at higher Reynolds number ($Re_\lambda = 2000 - 10^4$, figure 7).

A first critical test of the present model would require atmospheric/astrophysical Reynolds numbers, that is $Re_\lambda \sim 10^5$, where there is about half an order of magnitude between the maximum dissipation rate predicted by the present model (black line in figure 7) and a $Re_\lambda^{1.3}$ power law extrapolation from the DNS data (gray line). This difference is due to the sublayers emerging within the significant shear layers. The existence of sublayers is supported by the observations of Falgarone et al. [44] made in a molecular cloud. Their high-resolution measurements inside a ~0.03 pc thick large-scale shear layer [7,45] reveal thin shear layers with thicknesses as small as 3 mpc, which corresponds to the spatial resolution of the measurement. Therefore, the actual thickness may be less. For this molecular cloud (Polaris Flare), the Reynolds number is estimated using the density, $n_H = 200$ cm$^{-3}$, and the dissipation rate per unit volume, $2 \cdot 10^{-25}$ erg cm$^{-3}$ s$^{-1}$, typical for molecular clouds [7] and the temperature, 25 K, and the velocity dispersion at the 30 pc scale, $U = 4.8$ km/s, reported for the Polaris Flare [45]. This yields $Re_\lambda \approx 1.2 \cdot 10^5$ when assuming isotropic turbulence relations. At these conditions, the present model returns a significant shear layer thickness of $4\lambda_T = 0.029$ pc, while the estimated sublayer thickness is $4\lambda_T^* = 0.7$ mpc. The satisfactory agreement with the observations (keeping in mind that the sublayers were under-resolved) suggests that the observed shear layers can indeed be interpreted as significant shear layers and sublayers. Note that the observed sublayers cannot be Kolmogorov scale dissipation sheets, which are ~$10\eta$ thick [22] corresponding to 0.11 mpc. Hence, they are far below the spatial resolution and leave no detectable signal in the measurement. A detailed assessment of sublayers remains technically challenging, unfortunately. However, their effect on the maximum dissipation rate may be more readily observed in future atmospheric experiments. The subsequent formation of sub-sublayers could be relevant to molecular clouds at $Re_\lambda \sim 10^7$ [30]. At these Reynolds numbers, there is one and a half orders of magnitude between the dissipation rate maxima predicted by the constant power law extrapolation and the model including sub-sublayers. This has considerable implications for the local heating and chemistry within molecular clouds [7]. The chemical composition of clouds may, therefore, offer opportunities to test the different predictions for the dissipation rate. Any direct observation of sub-sublayers at those conditions seems unachievable with the current techniques.

Out of curiosity, the present model is extended to approach infinite Reynolds number. By repeating the processes described in §§2 and 3, additional generations of substructure appear when certain $Re_\lambda$ thresholds are exceeded. This affects the exponent of the local $Re_\lambda^\beta$ scaling for the maximum dissipation rate. It can be shown that the exponent changes according to $\beta = 1 + \sum_{m=1}^{n} \left(\frac{1}{2}\right)^m$, where $n$ is the substructure generation with $n = 1$ for the significant shear layers, $n = 2$ for the sublayers, $n = 3$ for the sub-sublayers and so on. This series converges to $\beta = 2$ for $n \to \infty$, which is the scaling predicted by multifractal theory. Indeed, the repeated process of creating sublayers within layers ultimately resembles a fractal. However, convergence is very slow. For $\beta = 1.9$, five generations are needed and $Re_\lambda \sim 10^{20}$, while $\beta = 1.95$ requires six generations and $Re_\lambda \sim 10^{40}$! Beyond these immense Reynolds numbers, constant power laws determine the dissipation properties and the limit as $Re_\lambda \to \infty$ may apply. Thus, in practice, ultimate-state turbulence does not exist from the perspective of the smallest scales. Consequently, the finite value of the Reynolds number remains significant in any currently imaginable flow.

As a final remark on the maximum dissipation rate, the single lognormal distribution returns a decreasing power law exponent as the Reynolds number increases (dashed grey line in figure 7). This suggests that substructures are indeed intrinsic to any explanation of the increasing power law exponent that would be observed in the data.

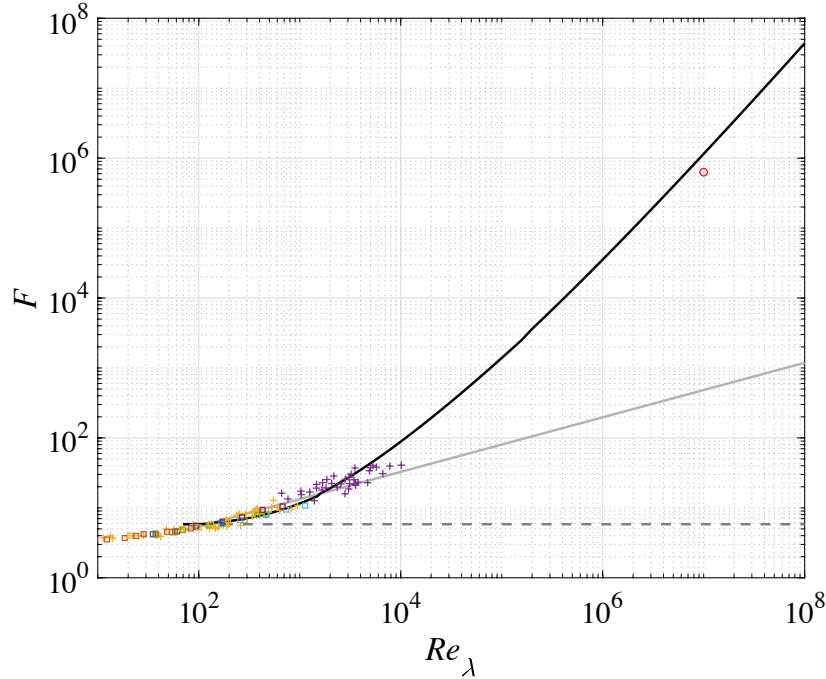

**Figure 8.** Flatness of the longitudinal velocity gradient predicted by the present model (black solid line) and a constant lognormal distribution (gray dashed line). The model results ($\alpha = 0.010$, $b = 0.67$) are compared with DNS ( ) and experimental data (+), which have been compiled by Sreenivasan & Antonia [17] and Ishihara et al. [15]. Consult these papers for reference to the original sources. Further note that the Reynolds numbers for the atmospheric experiments (purple +) have been adjusted from the reported values in [17] to the isotropic definition of $Re_\lambda$ as explained in the text. The grey solid line shows the power law proposed by Gylfason et al. [46]. The red circle indicates $F$ obtained from the smoothened model PDF at $Re_\lambda = 10^7$ (red dashed line in figure 6).

From the model PDFs (figure 5), the higher order moments of the dissipation rate can also be obtained. Here, we consider the model predictions for $\langle \varepsilon^2 \rangle$, which are compared to experimental and numerical data for the flatness, $F$, of the longitudinal velocity gradient over a wide range of Reynolds numbers. Assuming isotropy, $\langle \varepsilon^2 \rangle$ and $F$ are related according to [19,20]:

$$F = \frac{\langle \left(\frac{\partial u}{\partial x}\right)^4 \rangle}{\langle \left(\frac{\partial u}{\partial x}\right)^2 \rangle^2} = \frac{15}{7} \frac{\langle \varepsilon^2 \rangle}{\langle \varepsilon \rangle^2} \tag{4.1}$$

Furthermore, the normalized dissipation variance is given by $(\frac{7}{15}F - 1)$. The flatness, $F$, hence $\langle \varepsilon^2 \rangle$, is observed to be Reynolds number dependent ([15,17] and figure 8). This is expected since the PDF of the dissipation rate depends on the Reynolds number. Furthermore, the experimental and numerical data suggest that *F* does not follow a constant-power-law scaling, and that the exponent is increasing with $Re_\lambda$.

It should be pointed out that the data presented in figure 8 includes atmospheric data, which is anisotropic. In this flow, $Re_\lambda$ evaluated along the streamwise direction, as in [17], is much larger than $Re_\lambda$ evaluated along the other directions. To circumvent this problem, we adopt the isotropic definition of $Re_\lambda$ (see §2), which does not depend on direction and yields an intermediate value for $Re_\lambda$. The original source of this atmospheric data [47] used a definition close to ours. Therefore, their $Re_\lambda$ is shown in figure 8.

The prediction of the flatness factor *F* by the present model lies within the scatter of the data. Furthermore, the model captures the trend in the experimental and numerical data (figure 8), meaning it also reveals a non-constant power law behavior, where the exponent increases with Reynolds number. The significant shear layers and their sublayers provide a mechanism by which small-scale intermittency, hence *F*, can continue to increase with Reynolds number. Differences with respect to a constant power law scaling, such as proposed by Gylfason et al. [46] (gray line in figure 8), are significant. At high atmospheric Reynolds numbers, $Re_\lambda \sim 10^5$, the dissipation variance ($\sim F$) predicted by the model is more than an order of magnitude larger than the constant power law. At $Re_\lambda \sim 10^7$ it is more than three orders of magnitude larger.

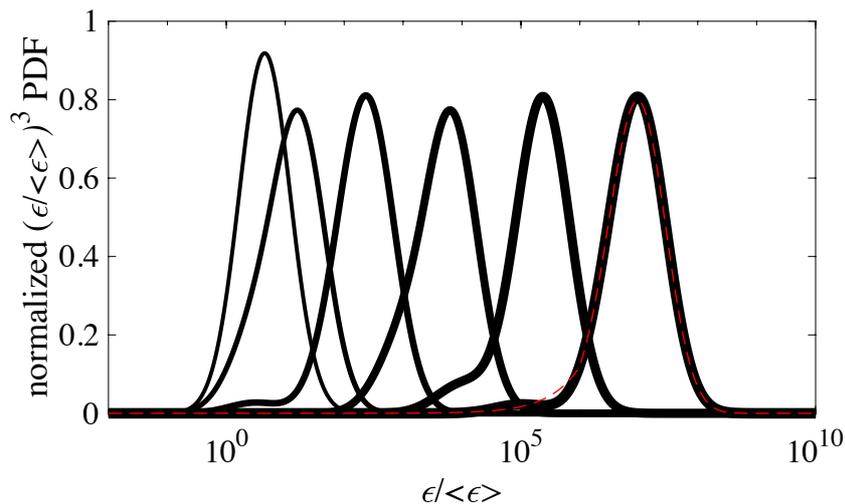

**Figure 9.** Flatness contributions normalized by *F*, such that the area under each graph is equal to one in this log-linear plot. Results are for $\alpha = 0.010$ and $b = 0.67$. For the legend refer to figure 6.

Figure 9 shows the dissipation ranges that contribute to *F* and hence to the dissipation variance. These ranges are found to overlap with the dissipation rates associated with the layers, sublayers and sub-sublayers (figure 4). Even though these layers occupy very little volume (figure 3), they contribute significantly to the variance. For higher order moments of the dissipation rate, the layer contributions will be even more important.

Furthermore, the local minima that exist in the model PDFs at very high Reynolds numbers (figure 6) have little effect on the magnitude of *F*. The model PDF with these minima removed (red dashed line) yields a distribution that is very close to the model PDF including the minima (figure 9). Removing the minima from the PDF reduces the flatness *F* by 45% at $Re_\lambda = 10^7$. The resulting *F* is marked in figure 8 by the red circle. However, the order of magnitude has not changed. Therefore, the Reynolds number dependence is not affected on the scale of the plot.

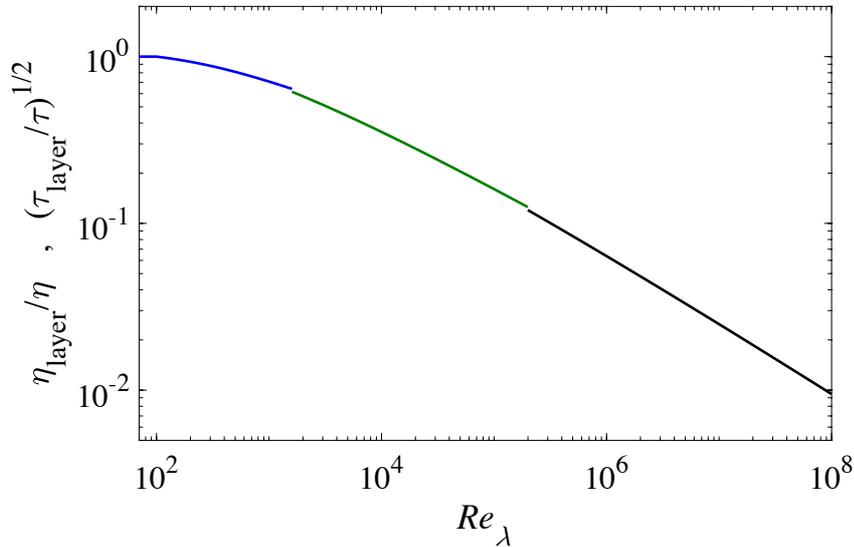

**Figure 10.** Local Kolmogorov length, $\eta_\text{layer}$, and time, $\tau_\text{layer}$, scales within the significant shear layers (blue), the sublayers (green) and sub-sublayers (black) normalized by global Kolmogorov length and time scale, $\eta$ and $\tau$ respectively. Results are for $\alpha = 0.010$ and $b = 0.67$.

## 5. The finest scales

The smallest length and time scales in a turbulent flow are generally considered to be proportional to the Kolmogorov scales, which are based on the global mean dissipation rate $\langle \varepsilon \rangle$. For example, the characteristic size of extreme dissipation structures is approximately $8\eta$ at low to moderate Reynolds number ($Re_\lambda < 1000$) [22].

However, in the present description of turbulence, there are thin layer regions where the local mean dissipation rate is enhanced significantly (figure 4). This affects the smallest scales. Local Kolmogorov scales can therefore be defined based on the local mean dissipation rate, as already shown in §2(b) for the local Kolmogorov length scale. Figure 10 presents the smallest local Kolmogorov length and time scales in the turbulent flow, $\eta_\text{layer}$ and $\tau_\text{layer}$ respectively. Depending on the Reynolds number, they occur in the significant shear layer, sublayer or sub-sublayer as indicated by the color coding in the figure. These smallest Kolmogorov scales determine the size of the smallest flow structures, e.g. the extreme dissipation structures are expected to be $8\eta_\text{layer}$ in size. As such, $\eta_\text{layer}$ and $\tau_\text{layer}$ determine the spatial and temporal requirements in a DNS and in experiments aiming to resolve small-scale turbulence, see [8,9] for a related discussion.

Up to $Re_\lambda = 1000$, the effect on the spatial scales is limited; $\eta_\text{layer}$ remains within 30% of the global Kolmogorov length scale $\eta$. Consequently, the implications for structure size and DNS spatial resolution are minor in this Reynolds number range. However, the smallest time scale has dropped appreciably to $\tau_\text{layer} = 0.50\tau$ at $Re_\lambda = 1000$. Time scales play an important role in the turbulent motion of inertial particles and droplets. For example, the dispersion, collision rate and growth of cloud drops are affected by the Stokes number, which includes the turbulent time scale. At typical cloud Reynolds numbers, $Re_\lambda \sim 10^4$ [48], we find $\tau_\text{layer} = 0.13\tau$, which means that locally the Stokes number changes by nearly an order of magnitude. This may result in quite different droplet dynamics and growth.

## 6. Conclusions

The main conclusion of this paper is that significant shear layers are intrinsic to any explanation and quantification of the dissipation rate statistics and extremes. Although intense dissipation has been observed previously within these layers, the significance of this observation for the dissipation scaling has not immediately been so clear, since these layers occupy very little volume. Their small volume fraction is, however, crucial for explaining the observed extremes of the dissipation rate.

The Reynolds number scaling of the maximum dissipation rate and the dissipation variance can be understood from a simple model, in which the turbulent flow is composed of large-scale background regions, significant shear layers, sublayers and sub-sublayer (figure 2). Furthermore, the significant shear layers are assumed to contribute a small, but constant, fraction to the overall dissipation. Since the volume fraction of the layers decreases with Reynolds number, the local mean dissipation rate within the layers increases. The maximum dissipation rate then follows the increase of the local mean.

Our fundamentally new result is that the model predicts a non-constant power law for the maximum dissipation rate and for the dissipation variance, with the power law exponent slowly increasing with the Reynolds number. The model results are in agreement with DNS data for homogenous isotropic turbulence up to at least $Re_\lambda = 1100$ (figures 7 and 8). By contrast, existing theories and data fits provide constant power laws, which do not accurately describe the trend of a slowly increasing exponent, as observed in the data. Moreover, the value of the exponent predicted by the existing theories is inconsistent with the data. The differences between the constant power law fits and the present model are significant at atmospheric and astrophysical Reynolds numbers as discussed in §4.

Furthermore, in our model, sublayers develop within the significant shear layers at large Reynolds numbers ($Re_\lambda > 1560$, §2). The sublayers further enhance intermittency and affect the Reynolds number scaling exponent for the maximum dissipation rate and the dissipation variance. The existence of sublayers is supported by observations in molecular clouds [44], but needs further confirmation from experiments or DNS. However, their predicted effect on the dissipation rate statistics may be more easily verified in measurements of atmospheric turbulence. At even higher Reynolds numbers ($Re_\lambda > 1.8 \cdot 10^5$) sub-sublayers are expected to appear within the sublayers.

Following the approach outlined in [42], the present dissipation PDFs may be used to estimate the PDFs of the velocity difference at a given separation. This allows evaluation of the velocity moments and structure functions, which could be an interesting extension of our work.

The present findings may be relevant to other non-classical properties of turbulence, such as the bottleneck in the turbulent kinetic energy spectrum, i.e. the spectral bump at wavenumbers of around $k\eta = 0.1$–$0.2$ (e.g. [11]). The 'bottleneck' has recently been associated with $1\lambda_T$–$4\lambda_T$ wide regions containing extreme dissipation [18]. These regions appear to be consistent with the significant shear layers in terms of their size and enhanced dissipation rate. Furthermore, the energy contained in the bump decreases with Reynolds number [11,49], which could be explained by the decreasing volume fraction of the significant shear layers (figure 3). Also, the bottleneck and the significant shear layers are both associated with significant interaction between disparate turbulent scales [28,50,51]. While other explanations cannot be ruled out, the above does suggest that significant shear layers contribute to the 'bottleneck'.

More generally, the significant shear layers provide a mechanism for direct large-scale to small-scale interactions with significant non-local energy transfer. Such interactions occur between the small scales within the significant shear layer and the large scales bounding them. As the Reynolds number increases, the large scales bound ever smaller scales (figure 10), which enhances the non-locality of the interaction in scale space. Kolmogorov's assumption of an inertial range, where the eddies are independent of the large scales and viscosity, breaks down near such layers, because all scales, including the smallest, are affected by the large energetic motions. Likewise, we expect that the large-scales will feel the effect of the very intense viscous dissipation at their bounds. This has implications for tracer particle dispersion at intermediate times, i.e. the so-called Richardson regime, which are explored elsewhere [52].

We end this paper by proposing a short addition to the famous Richardson rhyme describing the cascade of turbulent motions [53]. Consistent with our findings, it emphasizes the intermittent nature of the dissipation and its connection with large-scale shear:

*'But here and there*
*The eddies shear*
*In wobbly layers,*
*Where the micro-whirls are generated*
*And spikes of energy are dissipated.'*

Data accessibility. Data are available in the electronic supplementary material.

Authors' contributions. G.E.E. developed the model. T.I. carried out the numerical simulations. G.E.E. and T.I. analysed the results. The study was conceived by G.E.E. and J.C.R.H. All authors read, wrote and approved the manuscript.

Competing interests. We declare we have no competing interests.

Funding. TI was supported in part by JSPS KAKENHI Grant Number 20H01948 and MEXT as "Program for Promoting Researches on the Supercomputer Fugaku" (Toward a unified view of the universe: from large scale structures to planets).

Acknowledgements. The authors thank E. Falgarone for useful discussions on the layers in molecular clouds. The computer resources offered under the category of JHPCN Joint Research Projects by Research Institute for Information Technology, Kyushu University, and the Information Technology Center, Nagoya University, were used.